\documentclass[conference]{IEEEtran}
\IEEEoverridecommandlockouts
\usepackage{amsmath,amssymb,amsfonts}
\usepackage{algorithmic}
\usepackage{graphicx}
\usepackage{textcomp}
\usepackage{float}
\usepackage{xcolor}
\usepackage{booktabs}
\usepackage[
backend=biber,
style=numeric,
sorting=none
]{biblatex}
\def\BibTeX{{\rm B\kern-.05em{\sc i\kern-.025em b}\kern-.08em
    T\kern-.1667em\lower.7ex\hbox{E}\kern-.125emX}}

\begin{document}

\title{AI Literacy for Community Colleges: \\Instructors' Perspectives on Scenario-Based and Interactive Approaches to Teaching AI\\

\thanks{This material is based upon work supported by the AI Research Institutes Program funded by the National Science Foundation under the AI Institute for Societal Decision Making (NSF AI-SDM), Award No. 2229881.}
}

\makeatletter
\newcommand{\linebreakand}{%
  \end{@IEEEauthorhalign}
  \hfill\mbox{}\par
  \mbox{}\hfill\begin{@IEEEauthorhalign}
}
\makeatother

\author{
  \IEEEauthorblockN{Aparna Maya Warrier}
  \IEEEauthorblockA{\textit{Computer Science} \\
    \textit{Carnegie Mellon University}\\
    Pittsburgh, USA \\
    aparnamw@andrew.cmu.edu}
  \and
  \IEEEauthorblockN{Arav Agarwal}
  \IEEEauthorblockA{\textit{Computer Science} \\
    \textit{Carnegie Mellon University}\\
    Pittsburgh, USA \\
    arava@andrew.cmu.edu}
  \and
  \IEEEauthorblockN{Jaromir Savelka}
  \IEEEauthorblockA{\textit{Computer Science} \\
    \textit{Carnegie Mellon University}\\
    Pittsburgh, USA \\
    jsavelka@andrew.cmu.edu}
  \linebreakand 
  \IEEEauthorblockN{Christopher A Bogart}
  \IEEEauthorblockA{\textit{Computer Science} \\
    \textit{Carnegie Mellon University}\\
    Pittsburgh, USA \\
    cbogart@andrew.cmu.edu}
  \and
  \IEEEauthorblockN{Heather Burte}
  \IEEEauthorblockA{\textit{Computer Science} \\
    \textit{Carnegie Mellon University}\\
    Pittsburgh, USA \\
    hburte@andrew.cmu.edu}
}

\maketitle

\begin{abstract}
This research category full paper investigates how community college instructors evaluate interactive, no-code AI literacy resources designed for non-STEM learners. As artificial intelligence becomes increasingly integrated into everyday technologies, AI literacy—the ability to evaluate AI systems, communicate with them, and understand their broader impacts—has emerged as a critical skill across disciplines. Yet effective, scalable approaches for teaching these concepts in higher education remain limited, particularly for students outside STEM fields.

To address this gap, we developed \textit{AI User}, an interactive online curriculum that introduces core AI concepts through scenario-based activities set in real-world contexts. This study presents findings from four focus groups with instructors who engaged with \textit{AI User} materials and participated in structured feedback activities. Thematic analysis revealed that instructors valued exploratory tasks that simulated real-world AI use cases and fostered experimentation, while also identifying challenges related to scaffolding, accessibility, and multi-modal support. A ranking task for instructional support materials showed a strong preference for interactive demonstrations over traditional educational materials like conceptual guides or lecture slides. 

These findings offer insights into instructor perspectives on making AI concepts more accessible and relevant for broad learner audiences. They also inform the design of AI literacy tools that align with diverse teaching contexts and support critical engagement with AI in higher education.

\end{abstract}


\section{Introduction}
Artificial Intelligence (AI) systems are increasingly embedded into everyday tools, from chat-bots like ChatGPT to recommendation algorithms and predictive analytics. As AI tools become more pervasive across industries and everyday life, students and professionals alike are expected not only to use these systems but also to critically assess their outputs and implications \cite{Ng2021ConceptualizingAL}. This set of competencies, collectively referred to as AI literacy, has been defined as the ability to evaluate AI technologies, communicate with AI, and understand their ethical, societal, and economic implications \cite{Long2020WhatIA}.

There is a broad recognition in recent research that increasingly positions AI literacy as a foundational 21st-century skill, comparable to reading, writing, and digital literacy, and essential not only for computer scientists but for learners across all disciplines and professions \cite{Ng2021ConceptualizingAL}. Despite this growing importance, AI literacy education in higher education remains sparse and inconsistent, particularly outside of science, technology, engineering, and mathematics (STEM) domains \cite{Southworth2023DevelopingAM}. Most existing offerings are limited to computer science departments and often rely on programming-intensive coursework or traditional lectures, making them inaccessible to students without prior technical expertise \cite{Kong2021EvaluationOA}.

Recent reviews of AI literacy efforts in higher and adult education found that combining conceptual knowledge with hands-on, experiential learning was more effective than relying on either approach alone \cite{Laupichler2022ArtificialIL}. However, scalable, non-technical tools that support this balance remain rare, particularly those that cater to higher education contexts.

To address these gaps, we developed \textit{AI User}, a web-based curriculum designed to teach AI concepts to non-STEM learners through scenario-based, interactive activities. Prioritizing accessibility and engagement, the activities integrate visual scaffolding, conversational interfaces, and real-world use cases to help learners build conceptual understanding of AI without programming.

While prior work has focused on designing AI literacy curricula, fewer studies have examined how instructors interpret, evaluate, and adapt these resources for their classrooms \cite{Kong2021EvaluationOA}. Research in human-computer interaction and education has demonstrated the value of instructor perspectives in shaping technology-enabled learning environments \cite{cumbo2022using, ornekouglu2024systematic}. Approaches such as Value-Sensitive Design emphasize the importance of aligning learning tools with the values, needs, and constraints of educators \cite{VanBrummelen2020EngagingTT}.

In this study, we investigate how community college instructors interpret and evaluate hands-on AI literacy resources. To surface these perspectives, we conducted focus groups with instructors who engaged directly with the AI User curriculum. The design of the focus group discussions was informed by Value-Sensitive Design principles, prompting instructors to reflect on aspects such as contextual relevance, learner engagement, and support for critical thinking when evaluating the curriculum. Specifically, we address the following research questions:

\renewcommand\labelitemi{}
\begin{itemize}
\item \textbf{RQ1:} What benefits do instructors perceive when using hands-on educational exercises to teach AI?
\item \textbf{RQ2:} What challenges do instructors perceive when teaching AI literacy?
\item \textbf{RQ3:} What instructional support materials do instructors prefer for integrating AI curricula into their courses?
\end{itemize}
To answer these questions, we conducted four online focus groups with community college instructors who engaged directly with the interactive activities from the \textit{AI User} course. Through a reflexive thematic analysis \cite{braun2024thematic, terry2017thematic} of instructor feedback, we offer instructor insights for creating accessible and engaging AI literacy resources that support a range of instructional contexts in higher education.

\section{Related Work}
Efforts to expand AI literacy in higher education face several persistent challenges, particularly when aiming to reach non-STEM learners. This section reviews prior work on the challenges of designing accessible AI learning experiences, the barriers to integrating AI literacy into existing curricula, and the importance of incorporating instructor perspectives to inform scalable, relevant and context-sensitive instructional approaches.

\subsection{Challenges in Designing Accessible AI Education}
As AI literacy gains recognition as a critical competency for learners across disciplines \cite{Ng2021ConceptualizingAL, Long2020WhatIA}, educational initiatives face persistent challenges in reaching broader, non-technical audiences. Many university-level courses remain anchored in programming-based instruction or traditional lectures, making them less accessible to students without prior computational backgrounds \cite{Southworth2023DevelopingAM, Kong2022EvaluatingAI}.

A scoping review of 30 initiatives aimed at promoting AI literacy in higher and adult education found that combining knowledge-transfer with hands-on, experiential learning was particularly effective in supporting conceptual understanding \cite{Laupichler2022ArtificialIL}. However, few existing tools support this balance in a way that is scalable, no-code, and conceptually broad.

Efforts to broaden access have often relied on entry points such as robotics kits or tools like Google’s Teachable Machine \cite{carney2020teachable, ng2023review}. While these are engaging and intuitive, they present limitations. Robotics activities require physical equipment and infrastructure, posing challenges for scalability. Tools like Teachable Machine tend to focus on perceptual AI (e.g., image recognition) and offer limited opportunities to explore more foundational AI concepts such as data quality, uncertainty, and model behavior.

\subsection{Challenges in Integrating AI into Existing Curricula}
In addition to content accessibility, instructors face structural challenges when attempting to integrate AI literacy into their courses. Many instructional materials are not designed to be modular or adaptable, limiting their usefulness across varied classroom settings and student backgrounds \cite{mcgrath2023university}. 

One of the primary barriers to expanding AI literacy in higher education is the limited availability of adaptable instructional resources. Research on AI education in K-12 contexts highlights several persistent challenges that are also evident in higher education settings, including insufficient teaching materials, inaccessible or outdated tools, and a lack of hands-on learning opportunities \cite{Li2024FromUN, lee2022preparing}. Instructors often struggle due to the absence of structured, scalable resources that can be easily integrated into existing curricula and tailored to non-STEM backgrounds \cite{choi2023influence, yim2024teachers}.

Our approach with \textit{AI User} aims to address these challenges by offering a structured curriculum of web-based, scenario-driven projects that allow learners to engage with AI systems without programming through interactive activities. These projects focus on various topics such as model behavior, data labeling, and performance evaluation through guided, visual interactive activities. Furthermore, the curriculum is designed to be modular, allowing instructors to adopt materials flexibly within their own courses and pedagogical goals.

\subsection{Instructor Perspectives in AI Literacy Education}
While much of the existing work in AI literacy has focused on curriculum development and student learning outcomes, relatively few studies have explored how instructors in higher education interpret and adapt AI literacy materials for use in their classrooms \cite{Kong2021EvaluationOA}. Yet instructors play a pivotal role in shaping how emerging technologies like AI are introduced, particularly in non-STEM courses where students may have limited technical backgrounds.

Research on AI literacy in K-12, particularly in middle school and high school contexts, has emphasized the importance of incorporating instructor perspectives into the design of educational materials \cite{VanBrummelen2020EngagingTT, gardner2022co, kong2024developing}. These studies suggest that instructors play a critical role not only in implementing curricula but also in co-creating effective learning experiences. Participatory and co-design approaches reinforce the value of involving educators in shaping both the content and structure of AI education tools \cite{cumbo2022using, ornekouglu2024systematic}.

Our study extends these insights to the context of higher education by gathering qualitative feedback from community college instructors. We examine how instructors engage with no-code, scenario-based AI literacy materials and what instructional goals, opportunities, and concerns emerge from their experiences. These perspectives surface important pedagogical and practical considerations that can inform the design of more accessible and contextually relevant AI education resources for non-STEM learners.

\section{Methodology}
This study involved a series of four online focus groups with community college instructors to gather structured feedback on interactive, scenario-based AI literacy materials. Each focus group session lasted approximately 3.5 hours and included 4-5 participants, for a total of 17 instructors. Before the sessions, participants completed an online survey that included key demographic information, such as their primary department affiliation, courses taught, and institutional location. During each of the sessions, participants engaged with one interactive, no-code project from the \textit{AI User} course. 

\subsection{Participants}
The participants in this study were instructors currently teaching in community colleges across the United States. We recruited them through direct email outreach to instructors previously engaged with our research group, followed by snowball sampling through professional networks. Eligibility criteria included current employment at a community college in the United States, interest in teaching the \textit{AI User} course, and availability for a 30-minute online survey and a 3.5-hour online focus group.

\begin{figure}[htbp]
    \centering
    \includegraphics[width=\linewidth]{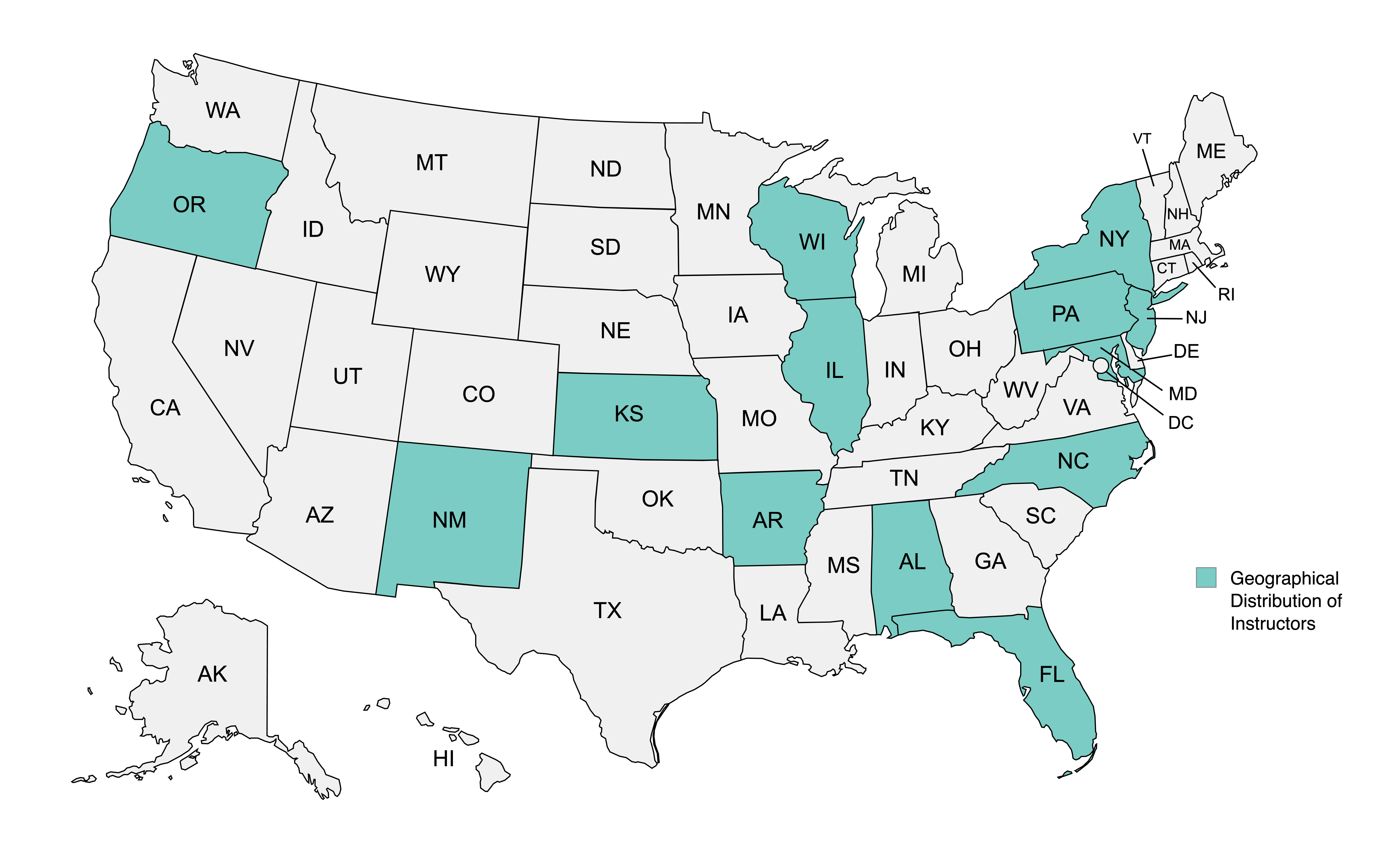}
    \caption{Geographic distribution of community college instructors who participated in the study.}
    \label{fig:map_distribution}
\end{figure}

\renewcommand{\arraystretch}{1.3}

\begin{table*}[htbp]
\centering
\caption{Participant Details Across Focus Group Sessions}
\begin{tabular}{lcll}
\toprule
\textbf{ID} & \textbf{Focus Group} & \textbf{Department} & \textbf{Courses Taught} \\
\midrule
P1  & Session 1 & Information Technology              & Intro to Computing, Business Analysis and Data Visualization \\
P2  & Session 1 & Information Technology              & Business Applications of AI, Intro to Analytics, Data Visualization \\
P3  & Session 1 & Business Hospitality and Technology & Intro to Programming, Creative Coding \\
P4  & Session 1 & Mathematics                         & Math Foundations (Algebra–Calculus), Python for Beginners \\
P5  & Session 2 & Computer Science                    & Computer Fundamentals, Intro to Web Development, Cybersecurity \\
P6  & Session 2 & Business and Information Systems    & Computer Ethics, Intro to Programming, Cybersecurity Fundamentals \\
P7  & Session 2 & Information Technology              & Intro to Programming \\
P8  & Session 2 & Computer Science                    & Computer Ethics, Cybersecurity, System Software \\
P9  & Session 3 & Computer Science                    & Python Programming, Data Analytics \\
P10 & Session 3 & Computer Science                    & Computer \& Informatics Science, Practical Programming \\
P11 & Session 3 & Computer Science                    & Computer Applications, Intro to Databases, Computer Maintenance \\
P12 & Session 3 & Computer Science                    & AI, Programming, Networking \\
P13 & Session 4 & Business and Information Systems    & Cybersecurity Fundamentals, Intro to Web Development \\
P14 & Session 4 & Engineering and Construction Technology & Intro to Programming \\
P15 & Session 4 & Applied Sciences and Technology     & Computer Applications, Intro to Programming \\
P16 & Session 4 & Computer Science                    & Digital Literacy, Computing Concepts \\
P17 & Session 4 & STEM                                & Robotics, Python, Statistics, Business Intelligence and Data Analytics, CAD \\
\bottomrule
\end{tabular}
\label{tab:participant_table}
\end{table*}

Based on survey responses, the seventeen instructors taught a range of courses, including STEM subjects like programming and cybersecurity, as well as interdisciplinary or non-STEM topics such as business, ethics, digital literacy, and creative coding. While most were not AI specialists, several had experience teaching related areas such as data literacy or digital ethics, aligning with the \textit{AI User} course’s focus on non-technical integration of AI. Fig. \ref{fig:map_distribution} provides an overview of the geographic distribution of community college instructors, spanning multiple regions in the United States. Table~\ref{tab:participant_table} presents additional details of the participants, including assigned identifiers (used in the text), attended focus group sessions, primary department affiliations, and courses taught. The availability of the participants guided the scheduling of the focus groups, each lasting 3.5 hours and facilitated by two researchers.

\subsection{Focus Group Design}
Each 3.5-hour focus group session was conducted via Zoom \cite{zoom} and followed a structured schedule (see Table~\ref{tab:focusgroupschedule}). The sessions included two design activities to elicit detailed instructor feedback.

First, a Rose-Bud-Thorn activity was used to gather reflections on one interactive, no-code project from the \textit{AI User} course. Instructors identified what worked well, what showed potential, and what needed improvement. Second, a ranking activity asked instructors to prioritize four types of instructional materials based on their usefulness for supporting classroom integration. These activities provided insights into both the perceived strengths of the course materials and the types of support instructors prefer when adopting new content. The following sections describe the purpose and design of each activity in more detail.

\begin{table}[htbp]
    \centering
    \caption{Schedule of Focus Groups}
    \begin{tabular}{ll}
        \toprule
        Time & Activity \\
        \midrule
        60 mins & Intro to AI User (Presentation) \\
        15 mins & Break \\
        60 mins & Rose, Bud, Thorn Activity \\
        15 mins & Break \\
        30 mins & Reflection on Rose, Bud, Thorn Activity (Discussion) \\
        15 mins & Ranking Educational Materials Activity \\
        15 mins & Reflection on Ranking Activity (Discussion) \\
        \bottomrule
    \end{tabular}
    \label{tab:focusgroupschedule}
\end{table}

\vspace{8pt}
\subsubsection{Rose, Bud, Thorn Activity}
Each focus group began with instructors engaging directly with a single project from the \textit{AI User} course. These projects were designed to teach foundational AI concepts through interactive, scenario-based activities that placed instructors in the role of learners. This direct experience served as a grounding activity, allowing participants to assess the pedagogical design and conceptual clarity of the material from both the learner's and the educator's perspective.

Following the project activity, instructors completed an individual written reflection using the Rose-Bud-Thorn framework. These reflections were documented using Miro \cite{miro}, an online collaborative white-boarding tool. This structured reflection method asked instructors to identify features of the activity that worked well in their context (``Roses''), elements that had potential for improvement or adaptation (``Buds''), and aspects that presented challenges, confusion, or limitations for implementation (``Thorns''). The Rose-Bud-Thorn framework encouraged instructors to critically evaluate the activity from the standpoint of classroom applicability, student engagement, and alignment with learning goals. After writing their reflections, the instructors participated in small group discussions in which they shared and elaborated on their Rose-Bud-Thorn responses. These conversations allowed participants to articulate their reasoning, respond to peer perspectives, and make comparisons between disciplinary and institutional contexts. In some cases, instructors revised or extended their reflections after these discussions, offering deeper insights into how their thinking evolved through peer interaction.

This iterative process, which began with hands-on participation, followed by structured individual reflection and peer discussion, enabled us to collect both individual and collective perspectives on each project. The Rose-Bud-Thorn framework provided a consistent structure across focus groups, allowing us to analyze patterns in instructor feedback, surface recurring strengths and concerns, and identify discipline-specific opportunities or barriers for implementing interactive AI literacy activities in non-STEM courses.

\subsubsection{Ranking Activity}
For the second design activity, the participants completed a ranking exercise aimed at evaluating supplemental educational materials designed to support \textit{AI User} projects. Instructors were offered four types of instructional support materials: interactive demos, concept guides, lecture slides, and reflection and discussion prompts. The interactive demos included hands-on visual tools such as Google's Teachable Machine \cite{carney2020teachable} and AI Explorables \cite{pairexplorables}, designed to help learners engage with AI concepts through experiential learning. The concept guides consisted of written and image-based materials---similar to online textbook content---that introduced key terms, explained foundational ideas, and provided real-world examples related to each AI topic. Lecture slides were developed to support in-class instruction and facilitate deeper student engagement through structured discussion. Reflection and discussion prompts encouraged students to reflect individually on the material and exchange ideas with peers in a collaborative setting.

This exercise began by providing participants with a clear and concise explanation of each supplementary resource, along with visual references to how they appeared in practice, ensuring a shared understanding of their intended use and potential educational impact.

The participants then individually ranked these resources based on their relevance and effectiveness. Individual rankings were subsequently discussed during each focus group, where participants explained the reasoning behind their preferences. These discussions provided additional context about how each type of instructional material aligned with their teaching styles, classroom settings, and student needs.

\subsection{Data Analysis}
Our dataset included video recordings of the co-design workshops, along with participant-generated artifacts such as written reflections from the Rose-Bud-Thorn activity and individual rankings of instructional support materials. All video recordings were transcribed for analysis. Two researchers conducted a reflexive thematic analysis \cite{braun2024thematic, terry2017thematic} of the written and verbal responses. We used an inductive approach, generating initial codes through repeated reading of the data. The researchers met iteratively to reflect on the coding process, discuss interpretations, and refine emerging themes. Consistent with reflexive thematic analysis, our focus was on developing a rich, interpretive understanding of participant perspectives rather than achieving coder agreement or applying a predefined codebook.

Instructor perspectives on the perceived benefits of interactive AI activities \textbf{(RQ1)} were primarily drawn from ``Rose'' responses and further elaborated during group discussions. Challenges and suggestions for improvement \textbf{(RQ2)} were identified through analysis of ``Thorn'' and ``Bud'' responses, contextualized by reflective comments shared during the sessions. Preferences for instructional support materials \textbf{(RQ3)} were assessed through a structured ranking activity. Rankings from 15 instructors were averaged using a 4-point scale, with higher values indicating stronger preference. A detailed summary of these results is provided in the following section.

\section{Results}

This section presents the key themes that emerged from our analysis of instructor feedback on the \textit{AI User} course. Insights are organized around the study’s three research questions and highlight both perceived strengths and challenges of using interactive, no-code AI literacy materials in higher education. In addition to qualitative feedback, this section also includes results from a ranking activity conducted during the focus groups to understand instructors’ preferences for instructional support materials. Together, the findings offer insight into how instructors perceive the design of AI education tools that are pedagogically meaningful, accessible to non-STEM learners, and adaptable to a range of instructional contexts.

\textbf{\\RQ1: What benefits do instructors perceive when using hands-on educational exercises to teach AI?}

\subsection{Visual Narratives to Support Early Engagement and Understanding}
One key theme that emerged from instructor feedback was the perceived value of using visual narratives and real-world scenarios to introduce AI concepts. In \textit{AI User}, each unit begins with a short visual scenario that presents a realistic application of AI, designed to provide context before the learners begin the interactive activities. These motivation tasks include examples such as analyzing sentiment in social media posts, preparing sensor data for predictive maintenance in aviation, building datasets for autonomous vehicles, and configuring drones for disaster response (see Fig. \ref{fig:motivation}). Several instructors suggested that these narrative elements may help make AI more approachable by connecting abstract ideas to familiar situations.

\begin{figure}[ht]
    \centering
    \includegraphics[width=\linewidth]{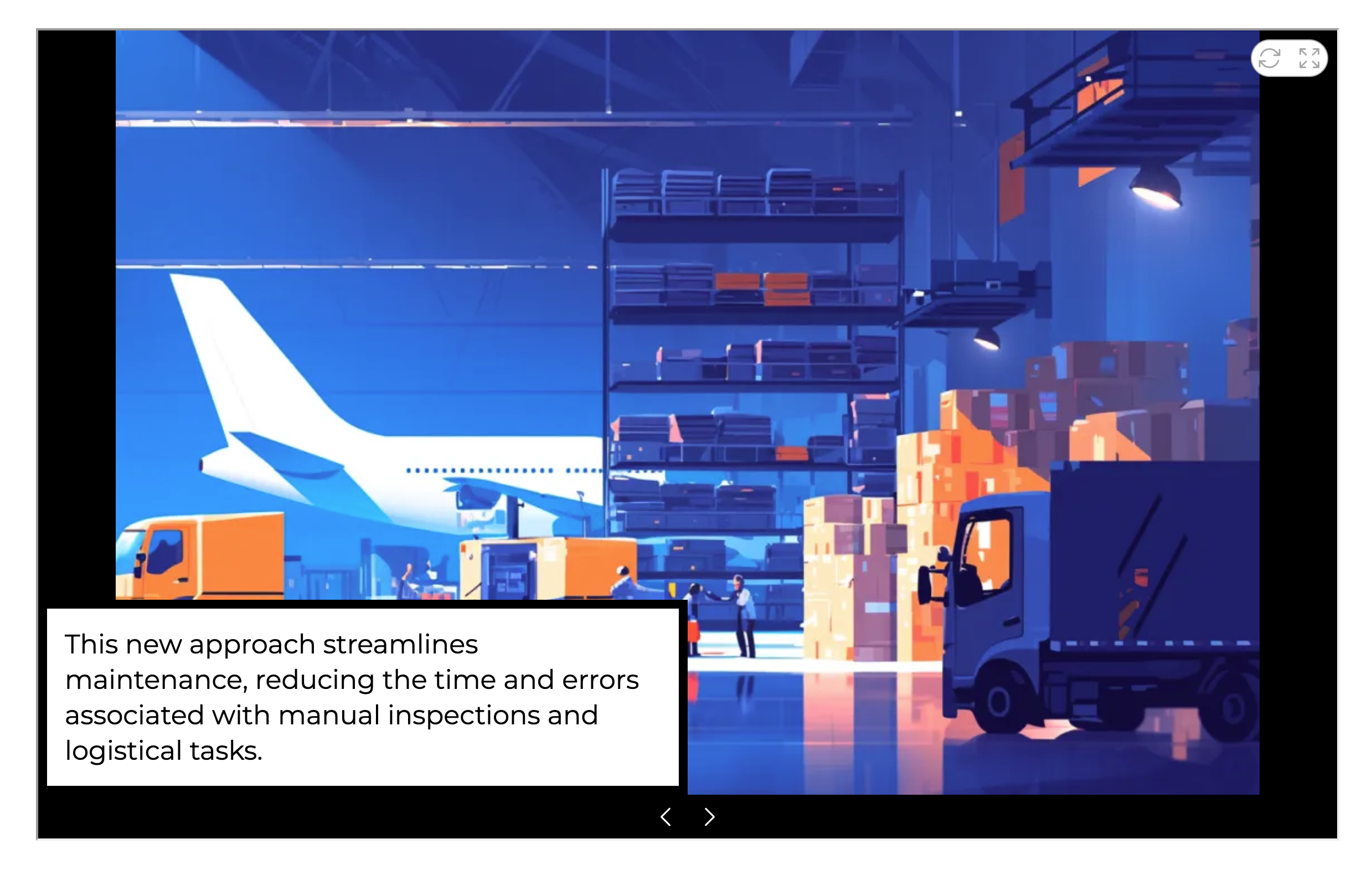}
    \caption{Snapshot of the motivation task introducing the predictive maintenance scenario in \textit{AI User}. The scenario presents a real-world context to learners in which an aviation company is exploring the use of AI to implement predictive maintenance for its airplanes.}
    \label{fig:motivation}
\end{figure}

This narrative framing was perceived as especially helpful in supporting early engagement, particularly for students who may not have a technical background. One participant noted that the interface was \textit{``immediately engaging with fun graphics''} (P2), and another commented that the \textit{``typical college scenario mentioned in social media posts is engaging''} (P1). These examples suggest that \textbf{relatable scenarios could serve as accessible entry points} into otherwise complex topics. Another instructor remarked,\textit{ ``The motivation provides just enough content and information to entice a person to want to learn more''} (P6), highlighting how the right level of detail may spark curiosity without overwhelming learners.

Instructors also commented on the realism of these narratives and how that could \textbf{potentially deepen learner interest}. One participant stated, \textit{``I liked having a believable real-life scenario for the exercise''} (P8), while another described the impact of the activity as \textit{``expanding a student's limits in terms of how they think about AI''} (P16). These comments suggest that grounding AI concepts in real-world contexts may help learners begin to understand the relevance of AI in different domains. The instructor even noted, \textit{``I learned something new myself''} (P16), pointing to the potential for these scenarios to be informative not only for students but also for educators unfamiliar with practical applications of AI.

\subsection{Simulated Environments and Animation to Support Explainability}
In \textit{AI User}, we use visual scaffolding and interactive animations to show how AI systems work, without requiring students to engage in programming or mathematical formulas. A recurring theme across instructor feedback was the potential for animated simulations and interactive environments to \textbf{make AI concepts more explainable and less abstract} for learners. Instructors indicated that this approach could help make concepts such as model performance, input-output relationships, and decision-making processes more concrete for their students, especially beginners.

In particular, several instructors perceived the animated model demonstrations as a useful way to support beginner understanding. In the first unit, for example, learners can pause and play a simulation to observe how a sentiment analysis model evaluates social media posts. One instructor described the animation as \textit{``outstanding''} in illustrating the tradeoff between speed and accuracy, adding that \textit{``you not only are told what the concept is but you experience the concept''} (P3). These animations were seen as a way to make terms like latency and accuracy more tangible, potentially helping learners grasp the tradeoffs in model performance. Instructors felt that being able to \textbf{observe these differences visually and in context}, rather than reading about them or encountering them through code, made these foundational AI concepts more accessible.

Instructors also highlighted the simulated environment's role-playing format as a potentially engaging feature for learners. Throughout the course, learners take on roles such as data annotator, analyst, or hardware specialist, completing tasks in a guided scenario. One instructor described the interface as \textit{``like a game''} (P10), and another shared, \textit{``This task works well and I think students would like the gamified appearance''} (P13). These perspectives suggest that \textbf{role-based simulations could encourage deeper learner involvement} by making AI tasks feel active and applied. One instructor stated, \textit{``We are often faced with students who require more than just a book, and putting this together with the visual aspects will definitely make it more user-friendly for students''} (P6), highlighting a broader concern about the limitations of traditional text-based learning formats.

\subsection{Interactive Activities for Iterative and Applied Learning}
Another theme that emerged from instructor feedback was the perceived value of interactive activities that support iterative learning, exploration, and applied decision making. In \textit{AI User}, learners simulate real-world AI tasks such as annotating sentiment in social media posts, creating image datasets for autonomous vehicles, and configuring drones for disaster response (see Fig. \ref{fig:interactive_task}. Instructors noted that these types of activities may help learners engage more deeply by applying knowledge, testing different approaches, and reflecting on outcomes.

Several instructors pointed to how the design of these activities may encourage learners to \textbf{experiment without fear of failure}. One participant commented, \textit{``Here the student can apply the knowledge learned. And even if the wrong answer is entered, students are still learning to improve as they go''} (P15). Another highlighted the opportunity to explore different outcomes of an activity, commenting that it was \textit{``great to explore multiple different configurations and their individual use cases''} (P16).

\begin{figure}[htbp]
    \centering
    \includegraphics[width=\linewidth]{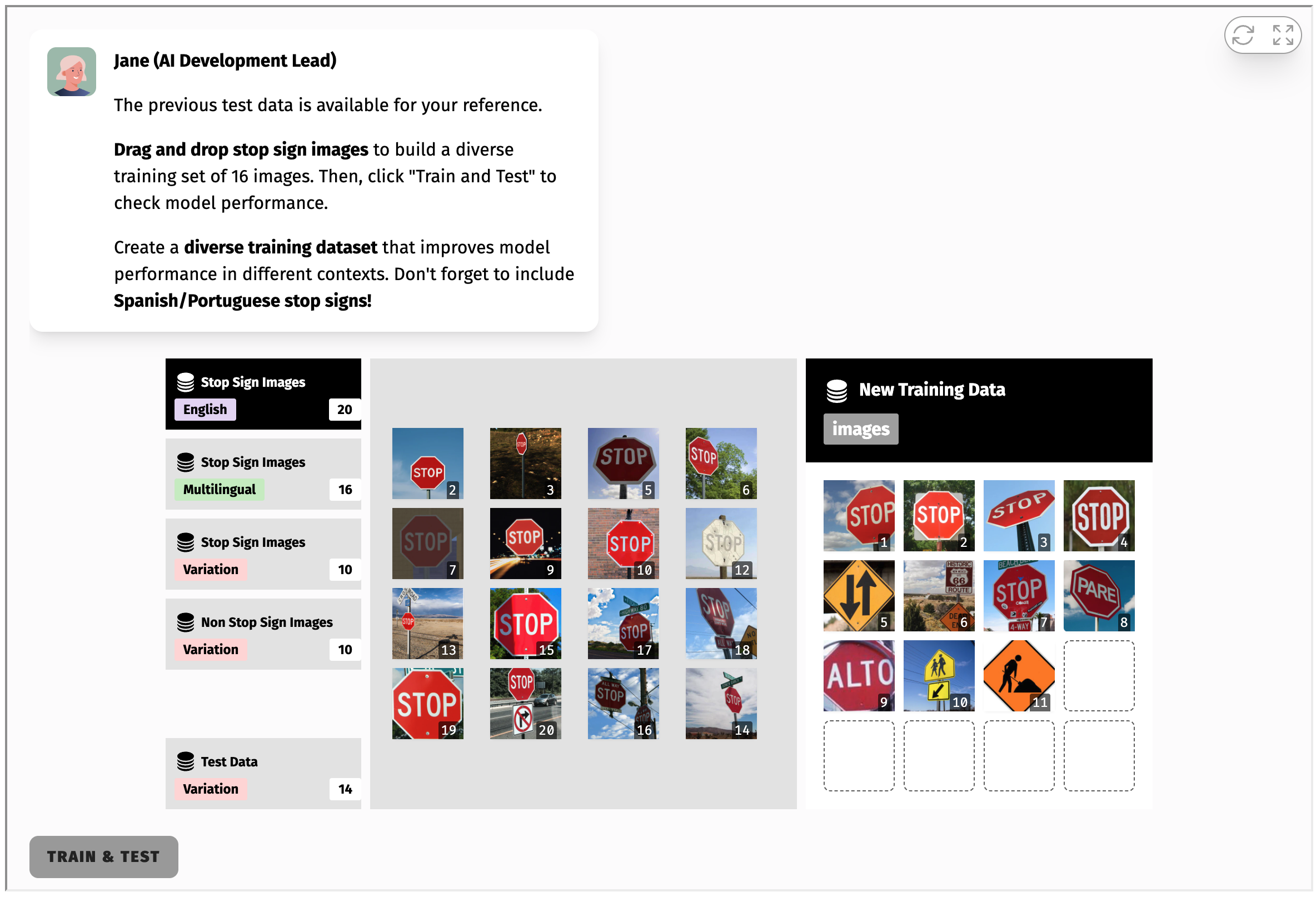}
    \caption{An interactive activity in \textit{AI User} where learners build an image dataset for autonomous vehicles that involves selecting and refining visual data to improve an AI system's ability to recognize stop signs.}
    \label{fig:interactive_task}
\end{figure}

The instructors also pointed out that interactive activities that replicate real-world tasks could \textbf{foster critical thinking and curiosity}. For example, in the sentiment labeling activity learners compare their annotations with those of simulated peers and supervisors. One instructor found it \textit{``clever''} (P2) that the simulated peer sometimes disagrees with the learner's choice, prompting reflection on how annotation is not always clear-cut. They remarked, \textit{``This is an excellent way to have students understand that AI at its core is still largely about patterns''} (P2).

These views suggest that instructors valued giving students the opportunity to \textbf{make their own decisions, observe consequences, and try again} if needed. One participant described the training data task for autonomous vehicles as \textit{``challenging, requires some decision-making''} (P10), while another noted that the drone configuration activity was \textit{``interactive, engaging and involved critical thinking''} (P15). This feedback points to the potential value of applying AI concepts through realistic tasks, which may support deeper understanding compared to more passive content delivery.

\subsection{Conversational Feedback and Accessible Language}
In \textit{AI User}, learners interact with simulated peers and supervisors through a chat-style interface. Instructors found the use of conversational tone and accessible language potentially helpful for \textbf{supporting student confidence}. Several instructors noted that this could make the experience more relatable and approachable, particularly for students without a technical background. One participant described it as \textit{``very relatable to the current college-aged student''} (P15), while another remarked, \textit{``I like the chat style delivery of the content''} (P17).

Many instructors noted that the overall tone of the feedback was positive and encouraging, which they felt may help reduce the pressure associated with making mistakes and encourage continued learner engagement. One instructor described it as a \textit{``nice positive, non-threatening tone describing corrections''} (P1), while another remarked, \textit{``incorrect is a chance for retraining, not a penalization''} (P2). This approach was seen as contributing to a more supportive learning experience, where learners may feel more willing to try again after an error. An instructor shared, \textit{``the feedback also explains why an answer is incorrect, if it is''} (P9), suggesting that these moments could reinforce reflection rather than discourage it.

In addition to tone and feedback structure, instructors emphasized the importance of \textbf{readability and clarity in instructional content}. Several instructors felt that avoiding complex terminology or technical jargon helped to keep learners engaged, particularly those without a background in AI. One participant noted, \textit{``This task keeps the verbiage at a level that most anyone can easily understand''} (P16), while another commented that it was \textit{``not overwhelming for those without a technical background''} (P15). These perspectives point to the role that tone and clarity may play in helping learners stay focused and build confidence when engaging with new or complex content.

\textbf{\\RQ2: What challenges do instructors perceive when teaching AI literacy?}

\subsection{Balancing Simplification and Conceptual Depth in Scaffolding AI Concepts}
One theme that emerged from instructor feedback was the challenge of balancing conceptual scaffolding with appropriate levels of complexity. Instructors noted that simplifying AI concepts too much may \textbf{limit student understanding}, while introducing too many technical terms or metrics at once could \textbf{overwhelm learners}, especially those new to the subject.

Several instructors pointed out that terms such as true positives, false negatives, or references to classification models might require more explanation or reinforcement. One participant suggested more elaboration for technical terms: \textit{``True Positive (TP), False Positive (FP) and so on—I can envision some students not understanding those terms''} (P10), and added that they had \textit{``copied the definitions to a file for reference.''} Another instructor suggested lightweight additions, such as a refresher or quick-link format, to help learners review unfamiliar concepts without interrupting the flow of the activity (P9).

At the same time, some instructors felt that the learning experience could benefit from more nuance, particularly when addressing model tradeoffs or decision metrics. For example, one participant reflected that \textit{``the tradeoffs aren't exactly black and white. I think some nuance in the example might add value''} (P7), while another commented that \textit{``the metrics such as cost shouldn't be too much of a giveaway''} (P2). These observations point to a tension in instructional design: how to simplify AI systems enough to promote accessibility, while retaining enough depth to reflect the complexity of real-world applications. Instructor feedback here underscores the need for flexible scaffolding strategies that can clarify foundational concepts without oversimplifying the learning experience.

\subsection{Balancing Exploration and Guidance in Interactive Activities}
Instructors noted that designing effective interactive tasks involves finding the right balance between exploration and structure. Several expressed concerns about learners bypassing conceptual learning through brute-force strategies, especially when activities resembled traditional formats like quizzes or multiple-choice questions. One instructor observed that \textit{``the labeling of the training posts is sometimes predictable''} (P4) and suggested randomizing elements to discourage solution sharing and encourage more authentic engagement. Instructor feedback indicated that activities may be \textbf{more effective when they mirror real-world AI tasks}, such as dataset creation or model configuration, rather than allowing learners to rely on trial-and-error to arrive at correct answers.

At the same time, instructors felt that overly open-ended activities, without adequate support, could lead to frustration. To address this, they suggested including mechanisms such as \textbf{progressively disclosed hints} or \textbf{opportunities to revisit decisions}. One participant proposed that \textit{``hints can be successively refined or revealed''} (P1), while another asked whether they can \textit{``retry without restarting the module?''} (P16). These features were viewed not as a way to simplify the task, but as potential supports for learners who may become stuck. This feedback suggests that carefully timed guidance could help maintain learner motivation while preserving the exploratory nature of the interactive activity.

\subsection{Supporting Accessibility and Multi-modal Engagement}
Instructor feedback emphasized the importance of making AI literacy content accessible and adaptable to a variety of student needs. Several instructors noted that dense or continuous blocks of information may be difficult for some students to navigate. One participant commented that while the task content was generally readable, \textit{``there may be some students who struggle to maintain that much information at once''} (P6). Another remarked, \textit{``my students often have short attention spans and breaking up content is helpful,''} (P5) pointing to the potential value of modular content delivery.

There was also interest in integrating multi-modal options such as \textbf{audio narration or text-to-speech features}. Instructors suggested that such features could support learners who may find extended text difficult to process or who face accessibility barriers. One participant proposed \textit{``possibly adding an audio option to hear the material as well,''} (P15) while others discussed the potential for incorporating voiceover or text-to-speech features to enable access for students with visual impairments. Instructors also expressed the view that many students today are accustomed to consuming information in short-form, audio-visual formats, and suggested that offering content in multiple modes could better align with how students typically engage with digital media.

\textbf{\\RQ3: What instructional support materials do instructors prefer for integrating AI curricula into their courses?}
\vspace{8pt}

To explore instructor preferences for instructional support materials, participants were asked to rank four content types: interactive demonstrations, lecture slides, conceptual guides, and reflection prompts. These rankings reflect how instructors anticipate using these materials in their own teaching practice. Rankings from 15 instructors were averaged using a scale where 4 represented the most preferred and 1 the least preferred option. Higher average ranks indicate stronger overall preference (see Table~\ref{tab:ranking_preferences}). Based on this, interactive demonstrations received the highest overall preference (highest mean rank = 3.20), followed by lecture slides (2.73), conceptual guides (2.47), and reflection prompts (1.60).

\begin{table}[htbp]
\centering
\caption{Average Rankings of Instructional Support Materials (N = 15)}
\begin{tabular}{l c c}
\toprule
\textbf{Content Type} & \textbf{Mean Rank} & \textbf{Ranked 1st (n)} \\
\midrule
Interactive Demos          & 3.20 & 7 \\
Lecture Slides             & 2.73 & 4 \\
Conceptual Guides          & 2.47 & 3 \\
Reflection Prompts         & 1.60 & 1 \\
\bottomrule
\end{tabular}
\label{tab:ranking_preferences}
\end{table}

Instructor feedback suggests that interactive demonstrations were seen as potentially effective in helping learners engage with AI concepts through observation and experimentation. Several instructors described these materials as useful for illustrating abstract ideas in a more applied, visual format, which they felt may support understanding for students without technical backgrounds.

Preferences for other materials appeared to reflect instructional context. Instructors teaching face-to-face courses tended to prefer lecture slides, which they felt could support structured explanations and in-class discussions. Those teaching online or asynchronously often favored conceptual guides, citing its flexibility for self-paced learning and independent review.

Reflection prompts were generally ranked lowest, as several instructors felt that structured online reflections often felt forced to students, and lacked the authenticity of organic, in-person classroom discussions. They emphasized that meaningful reflection tends to emerge through spontaneous dialogue, debate, or discussion among students. However, several instructors noted that if integrated directly into the interactive activities, reflection prompts could effectively reinforce key concepts and promote critical thinking during the learning process.

These findings suggest that while preferences vary by instructional context, instructors tend to prioritize materials that are interactive, adaptable, and conducive to active learning. Resources that support both conceptual clarity and student autonomy were viewed as particularly valuable for introducing AI in accessible and engaging ways.

\section{Discussion}

Instructor feedback from this study offers important insight into how hands-on AI literacy tools can be designed to better support learners in non-STEM domains. While previous research has identified gaps in AI education for general audiences \cite{Southworth2023DevelopingAM, Kong2021EvaluationOA}, these findings provide grounded perspectives into what educators perceive as both effective and challenging when implementing hands-on AI learning in classrooms.

One key insight is that instructors appeared to value the interactive, no-code activities that provide conceptual clarity without oversimplification. This supports previous work that has critiqued AI courses for being either too programming-intensive or overly passive in delivery \cite{Laupichler2022ArtificialIL}. Instructors in this study described interactive tasks as a potentially productive middle ground for beginners. These activities were perceived as allowing learners to explore, make decisions, and revise their understanding, while also remaining accessible to those without technical backgrounds. The instructor feedback adds support to ongoing efforts to build hands-on, simulation-based environments that emphasize intuitive learning and critical thinking rather than coding fluency.

Furthermore, the results highlight the perceived value of embedding technical content within familiar, real-world scenarios. Instructors described narrative-based introductions as a helpful way to engage learners and frame the purpose of each task. This aligns with prior findings on the importance of contextual learning \cite{Kong2022EvaluatingAI} and suggests that scenario-driven entry points may help learners make meaningful connections between abstract AI concepts and practical applications.

At the same time, the study adds nuance to the idea of hands-on learning by highlighting the importance of balance. While exploration and open-ended problem solving were valued, instructors also identified the need for structured guidance to avoid learner frustration. This reinforces findings in related work on the importance of instructional scaffolding in self-guided learning environments \cite{Laupichler2022ArtificialIL, ng2023review}, but expands on it by pointing to specific mechanisms, such as progressive hints and opportunities to revise decisions, that may support persistence without undermining exploration.

Instructor feedback also surfaces gaps in current research on inclusive AI education. While most hands-on AI tools rely heavily on visual interaction, there is limited research on how to design these tools for learners with disabilities or impairments. Instructor suggestions for audio narration and modular content reflect a growing recognition that inclusive design will be essential for scaling AI literacy efforts. As web-based learning tools for AI literacy continue to grow in popularity, ensuring they support diverse modes of interaction---including screen readers, voice input, and alternative formats---may significantly expand their reach.

Overall, the findings suggest that instructors are seeking educational tools that balance structure with flexibility, support active learning without being overly complex, and prioritize inclusive design. While prior research has identified general principles for AI literacy, instructor perspectives in this study offer more detailed insight into what implementation may look like in practice for higher education. By centering instructor experience, this study contributes to a growing body of work that aims to develop AI education strategies that are both pedagogically effective and widely accessible.

\section{Limitations and Future Work}

This study was based on structured qualitative feedback from instructors who reviewed the \textit{AI User} course and its educational content, but did not involve direct observation of student use. While instructor perspectives offer valuable insight into pedagogical relevance and perceived effectiveness, they do not capture how students actually experience and engage with the materials. Future work will address this gap by conducting pilot studies with students, allowing for comparison between instructor expectations and student outcomes. Such studies will also provide opportunities to gather quantitative data on learner interactions, engagement, and performance within the platform.

In addition, several interactive activities will be redesigned based on the feedback gathered in this study. These revisions aim to better align the content and interface with instructors’ instructional goals, particularly around scaffolding, exploration, and accessibility. Piloting these redesigned activities with student users will allow for iterative evaluation and refinement, contributing to the development of AI literacy tools that are both educationally effective and inclusive. Continued focus groups with instructors will also be important to assess the effectiveness of changes and gather ongoing input to ensure the learning design remains responsive to classroom needs.

\section{Conclusion}

This study examined community college instructors’ perspectives on the use of hands-on, scenario-based tools for teaching AI literacy to non-STEM learners. Through structured feedback and qualitative analysis, the findings highlight instructional strategies that instructors found effective, including narrative framing, interactive simulations, and conversational feedback. They also point to key challenges, such as balancing guidance with open-ended exploration, supporting accessibility, and scaffolding technical concepts in ways that remain accurate without becoming overly simplified.

The results contribute to ongoing conversations about broadening access to AI education by offering grounded insights into what makes instructional tools usable and pedagogically sound from the perspective of educators. In particular, the study highlights a preference for interactive, no-code approaches that promote engagement and critical thinking, while also suggesting the need for inclusive design that accommodates a broad range of learners and teaching contexts.

As AI literacy efforts continue to expand across disciplines and institutions, incorporating instructor feedback into the design of educational tools will be essential. Future work should explore how these tools perform in real classroom settings, how students experience them across different demographics, and how design strategies can be further refined to support equity, accessibility, and meaningful learning at scale.

\section{Acknowledgments}

We sincerely thank the community college instructors who participated in this research for generously sharing their time, insights, and feedback. We are grateful to Hanzhi Yin for developing the interactive course activities used in the study. We also thank the research assistants who contributed to the design, development and testing of these activities.

\printbibliography

@article{Long2020WhatIA,
  title={What is AI Literacy? Competencies and Design Considerations},
  author={Duri Long and Brian Magerko},
  journal={Proceedings of the 2020 CHI Conference on Human Factors in Computing Systems},
  year={2020},
  url={https://api.semanticscholar.org/CorpusID:211264278}
}

@article{Ng2021ConceptualizingAL,
  title={Conceptualizing AI literacy: An exploratory review},
  author={Davy Tsz Kit Ng and Jac Ka Lok Leung and Samuel Kai-Wah Chu and Maggie Shen Qiao},
  journal={Comput. Educ. Artif. Intell.},
  year={2021},
  volume={2},
  pages={100041},
  url={https://api.semanticscholar.org/CorpusID:244514711}
}

@article{Southworth2023DevelopingAM,
  title={Developing a model for AI Across the curriculum: Transforming the higher education landscape via innovation in AI literacy},
  author={Joanna R. Southworth and Kati Migliaccio and Joe Glover and Janet M. Glover and David Reed and Christopher McCarty and Joel H. Brendemuhl and Aaron O. Thomas},
  journal={Comput. Educ. Artif. Intell.},
  year={2023},
  volume={4},
  pages={100127},
  url={https://api.semanticscholar.org/CorpusID:256159931}
}

@article{Kong2021EvaluationOA,
  title={Evaluation of an artificial intelligence literacy course for university students with diverse study backgrounds},
  author={Siu Cheung Kong and William Man Yin Cheung and Guo Zhang},
  journal={Comput. Educ. Artif. Intell.},
  year={2021},
  volume={2},
  pages={100026},
  url={https://api.semanticscholar.org/CorpusID:236678136}
}

@article{Li2024FromUN,
  title={``From Unseen Needs to Classroom Solutions": Exploring AI Literacy Challenges \& Opportunities with Project-based Learning Toolkit in K-12 Education},
  author={Hanqi Li and Ruiwei Xiao and Hsuan Nieu and Ying-Jui Tseng and Guanze Liao},
  journal={ArXiv},
  year={2024},
  volume={abs/2412.17243},
  url={https://api.semanticscholar.org/CorpusID:274982661}
}

@article{Laupichler2022ArtificialIL,
  title={Artificial intelligence literacy in higher and adult education: A scoping literature review},
  author={Matthias Carl Laupichler and Alexandra Aster and Jana Schirch and Tobias Raupach},
  journal={Comput. Educ. Artif. Intell.},
  year={2022},
  volume={3},
  pages={100101},
  url={https://api.semanticscholar.org/CorpusID:252551067}
}

@article{Kong2022EvaluatingAI,
  title={Evaluating artificial intelligence literacy courses for fostering conceptual learning, literacy and empowerment in university students: Refocusing to conceptual building},
  author={Siu Cheung Kong and William Man Yin Cheung and Guo Zhang},
  journal={Computers in Human Behavior Reports},
  year={2022},
  url={https://api.semanticscholar.org/CorpusID:251238054}
}

@article{VanBrummelen2020EngagingTT,
  title={Engaging Teachers to Co-Design Integrated AI Curriculum for K-12 Classrooms},
  author={Jessica Van Brummelen and Phoebe Lin},
  journal={Proceedings of the 2021 CHI Conference on Human Factors in Computing Systems},
  year={2020},
  url={https://api.semanticscholar.org/CorpusID:221856678}
}

@article{ng2023review,
  title={A review of AI teaching and learning from 2000 to 2020},
  author={Ng, Davy Tsz Kit and Lee, Min and Tan, Roy Jun Yi and Hu, Xiao and Downie, J Stephen and Chu, Samuel Kai Wah},
  journal={Education and Information Technologies},
  volume={28},
  number={7},
  pages={8445--8501},
  year={2023},
  publisher={Springer}
}

@article{mcgrath2023university,
  title={University teachers' perceptions of responsibility and artificial intelligence in higher education-An experimental philosophical study},
  author={McGrath, Cormac and Pargman, Teresa Cerratto and Juth, Niklas and Palmgren, Per J},
  journal={Computers and Education: Artificial Intelligence},
  volume={4},
  pages={100139},
  year={2023},
  publisher={Elsevier}
}

@inproceedings{gardner2022co,
  title={Co-designing an AI curriculum with university researchers and middle school teachers},
  author={Gardner-McCune, Christina and Touretzky, David and Cox, Bryan and Uchidiuno, Judith and Jimenez, Yerika and Bentley, Betia and Hanna, William and Jones, Amber},
  booktitle={Proceedings of the 54th ACM Technical Symposium on Computer Science Education V. 2},
  pages={1306--1306},
  year={2022}
}

@inproceedings{lee2022preparing,
  title={Preparing high school teachers to integrate AI methods into STEM classrooms},
  author={Lee, Irene and Perret, Beatriz},
  booktitle={Proceedings of the AAAI conference on artificial intelligence},
  volume={36},
  number={11},
  pages={12783--12791},
  year={2022}
}

@article{choi2023influence,
  title={Influence of pedagogical beliefs and perceived trust on teachers’ acceptance of educational artificial intelligence tools},
  author={Choi, Seongyune and Jang, Yeonju and Kim, Hyeoncheol},
  journal={International Journal of Human--Computer Interaction},
  volume={39},
  number={4},
  pages={910--922},
  year={2023},
  publisher={Taylor \& Francis}
}

@article{yim2024teachers,
  title={Teachers' perceptions, attitudes, and acceptance of artificial intelligence (AI) educational learning tools: An exploratory study on AI literacy for young students},
  author={Yim, Iris Heung Yue and Wegerif, Rupert},
  journal={Future in Educational Research},
  volume={2},
  number={4},
  pages={318--345},
  year={2024},
  publisher={Wiley Online Library}
}

@inproceedings{carney2020teachable,
  title={Teachable machine: Approachable Web-based tool for exploring machine learning classification},
  author={Carney, Michelle and Webster, Barron and Alvarado, Irene and Phillips, Kyle and Howell, Noura and Griffith, Jordan and Jongejan, Jonas and Pitaru, Amit and Chen, Alexander},
  booktitle={Extended abstracts of the 2020 CHI conference on human factors in computing systems},
  pages={1--8},
  year={2020}
}

@incollection{braun2024thematic,
  title={Thematic analysis},
  author={Braun, Virginia and Clarke, Victoria},
  booktitle={Encyclopedia of quality of life and well-being research},
  pages={7187--7193},
  year={2024},
  publisher={Springer}
}

@article{terry2017thematic,
  title={Thematic analysis},
  author={Terry, Gareth and Hayfield, Nikki and Clarke, Victoria and Braun, Virginia and others},
  journal={The SAGE handbook of qualitative research in psychology},
  volume={2},
  number={17-37},
  pages={25},
  year={2017},
  publisher={SAGE Publications Ltd}
}

@article{ornekouglu2024systematic,
  title={A systematic literature review on co-design education and preparing future designers for their role in co-design},
  author={{\"O}rneko{\u{g}}lu-Sel{\c{c}}uk, Melis and Emmanouil, Marina and Hasirci, Deniz and Grizioti, Marianthi and Van Langenhove, Lieva},
  journal={CoDesign},
  volume={20},
  number={2},
  pages={351--366},
  year={2024},
  publisher={Taylor \& Francis}
}

@article{cumbo2022using,
  title={Using participatory design approaches in educational research},
  author={Cumbo, Bronwyn and Selwyn, Neil},
  journal={International Journal of Research \& Method in Education},
  volume={45},
  number={1},
  pages={60--72},
  year={2022},
  publisher={Taylor \& Francis}
}

@article{kong2024developing,
  title={Developing an artificial intelligence literacy framework: Evaluation of a literacy course for senior secondary students using a project-based learning approach},
  author={Kong, Siu-Cheung and Cheung, Man-Yin William and Tsang, Olson},
  journal={Computers and Education: Artificial Intelligence},
  volume={6},
  pages={100214},
  year={2024},
  publisher={Elsevier}
}

@misc{pairexplorables,
  author       = {{People \& AI Research (PAIR), Google}},
  title        = {AI Explorables},
  howpublished ={\url{https://pair.withgoogle.com/explorables/}},
  note         = {Accessed: 2025-04-10}
}

@misc{miro,
  author       = {{Miro}},
  title        = {Miro: Online Collaborative Whiteboard Platform},
  howpublished = {\url{https://miro.com/}},
  note         = {Accessed: 2025-04-10}
}

@misc{zoom,
  author       = {{Zoom Video Communications, Inc.}},
  title        = {Zoom},
  howpublished = {\url{https://zoom.us}},
  note         = {Accessed: 2025-04-10}
}

\end{document}